\documentclass[10pt,twocolumn,showpacs,superscriptaddress,amsmath,amssymb,osajnl,floatfix]{revtex4-1}

\usepackage{graphicx}
\usepackage{dcolumn}
\usepackage{bm}
\usepackage{color}
\usepackage[colorlinks, linkcolor=blue, anchorcolor=blue,citecolor=blue]{hyperref}
\usepackage{amsmath}

\begin{document}
	\author{Feng Wan}  \affiliation{MOE Key Laboratory for Nonequilibrium Synthesis and Modulation of Condensed Matter, School of Science, Xi'an Jiaotong University, Xi'an 710049, China}
	\author{Kun Xue}	\affiliation{MOE Key Laboratory for Nonequilibrium Synthesis and Modulation of Condensed Matter, School of Science, Xi'an Jiaotong University, Xi'an 710049, China}
	\author{Zhen-Ke Dou}	\affiliation{MOE Key Laboratory for Nonequilibrium Synthesis and Modulation of Condensed Matter, School of Science, Xi'an Jiaotong University, Xi'an 710049, China}
		\author{Karen Z. Hatsagortsyan}
	\affiliation{Max-Planck-Institut f\"{u}r Kernphysik, Saupfercheckweg 1,
		69117 Heidelberg, Germany}	
	\author{Wenchao Yan}\email{Wenchao.Yan@eli-beams.eu}	\affiliation{Institute of Physics ASCR, v.v.i. (FZU), ELI BEAMLINES, Za Radnic\'i 835, Doln\'i B\v re\v zany, 252241, Czech Republic} 	
		\author{Danila Khikhlukha}	\affiliation{Institute of Physics ASCR, v.v.i. (FZU), ELI BEAMLINES, Za Radnic\'i 835, Doln\'i B\v re\v zany, 252241, Czech Republic}
	\author{Sergei V. Bulanov}	\affiliation{Institute of Physics ASCR, v.v.i. (FZU), ELI BEAMLINES, Za Radnic\'i 835, Doln\'i B\v re\v zany, 252241, Czech Republic} \affiliation{Kansai Photon Science Institute, National Institutes for Quantum and Radiological Science and Technology, 8-1-7 Umemidai, Kizugawa-shi, Kyoto, 619-0215, Japan} \affiliation{Prokhorov Institute of General Physics of the Russian Academy of Sciences, Vavilov St. 38, 119991, Moscow, Russia}
		\author{Georg Korn}	\affiliation{Institute of Physics ASCR, v.v.i. (FZU), ELI BEAMLINES, Za Radnic\'i 835, Doln\'i B\v re\v zany, 252241, Czech Republic} 	
	\author{Yong-Tao Zhao}	\affiliation{MOE Key Laboratory for Nonequilibrium Synthesis and Modulation of Condensed Matter, School of Science, Xi'an Jiaotong University, Xi'an 710049, China}
	\author{Zhong-Feng Xu}	\affiliation{MOE Key Laboratory for Nonequilibrium Synthesis and Modulation of Condensed Matter, School of Science, Xi'an Jiaotong University, Xi'an 710049, China}
	
	\author{Jian-Xing Li}\email{jianxing@xjtu.edu.cn}	\affiliation{MOE Key Laboratory for Nonequilibrium Synthesis and Modulation of Condensed Matter, School of Science, Xi'an Jiaotong University, Xi'an 710049, China}
	\bibliographystyle{apsrev4-1}
	
	\title{Imprint of the stochastic nature of photon emissions by electrons on the proton energy spectra in the laser-plasma interaction}
	
	\date{\today}

\begin{abstract}

The impact of stochasticity effects (SEs) in photon emissions on the proton energy spectra during laser-plasma interaction is theoretically investigated in the quantum radiation-dominated regime, which may facilitate SEs experimental observation. We calculate the photon emissions quantum mechanically and the plasma dynamics semiclassically via two-dimensional particle-in-cell simulations. An ultrarelativistic plasma generated and driven  by an ultraintense laser pulse head-on collides  with another strong laser pulse, which decelerates the electrons due to radiation-reaction effect and results in a significant compression of the proton energy spectra  because of the charge separation force.  In the considered regime the SEs are demonstrated in the shift of the mean energy of the protons up to hundreds of MeV. This  effect is robust with respect to the laser and target parameters and measurable in soon  available strong laser facilities.

\end{abstract}

\maketitle
\section{Introduction}

With the rapid development of ultrashort ultraintense laser techniques, the laser peak intensity  can reach up to $10^{22}$ W/cm$^2$ at present and   is expected to attain $10^{24}-10^{26}$ W/cm$^2$ in the facilities under construction  \cite{Vulcan,ELI,Exawatt,Yanovsky2008,Danson_2015_Petawatt, Sung_2017_42,Mourou_2011_Extreme,Zou_2015_Design}, which can
be exploited to gain new insight for fundamental physics \cite{Piazza_2012_Extremely, Macchi_2013_Ion, Bulanov_2009_Relativistic}. In ultrastrong laser fields, the electron dynamics is ultrarelativistic and characterized by multiphoton nature \cite{Nikishov_1964_Quantum, Ritus_1985_Quantum}, for instance, recently demonstrated experimentally in multiphoton Thomson scattering \cite{Yan_2017}. In this situation radiation reaction (RR) effects \cite{Abraham_1905,Lorentz_1909,Dirac_1938,Heitler_1941} can be conspicuous. In particular, when
the energy loss of an electron due to radiation is comparable with its own energy, the electron dynamics is significantly modified due to RR, bringing about the radiation dominated regime \cite{Koga2005,Piazza_2012_Extremely}.
In strong fields quantum effects in radiation take place when the photon emission recoil is discernible, i.e. when the emitted photon energy is comparable with the electron one. In the quantum radiation regime, the photon emissions are discrete and probabilistic, which combined with the large recoil lead to
stochasticity effects (SEs) in electron dynamics
\cite{Shen_1972_Energy, Tamburini_2014_Electron, Wang_2015_Signatures}.
SEs can quantitatively enhance the hard-photon emission in ultrastrong field due to, so-called, electron straggling effect \cite{Shen_1972_Energy,Blackburn_2014_Quantum},
quench radiation losses in subcycle PW lasers \cite{Harvey_2017_Quantum}, heat electron beams in plasma \cite{Mendonca_1983, Sheng_2002} and in vacuum \cite{Neitz_2013_Stochasticity, Neitz_2014_Electron, Bashinov_2015_Impact}, and reshape the angular spectra of electron beams \cite{Li_2018_Electron} and emitted photons  \cite{Li_2017_Angle, Li_2014_Robust}.

Recently, experimental evidence for quantum RR effects has been demonstrated in the radiation spectra of an ultrarelativistic positron beam traversing a silicon slab \cite{Wistisen_2018_Experimental} and in the energy loss of an electron beam after colliding with a strong laser pulse \cite{Poder_2018_Experimental,Cole_2018_Experimental}. However, in those experiments all quantum properties, including SEs and photon recoil effects, are entangled.
Thus, the  detection of the sole signature of SEs is still a challenging and opening question.

 Although RR directly disturbs the electron dynamics during laser-plasma interaction, it has an indirect influence on the ion dynamics due to the charge separation quasi-static fields, which are modified by RR disturbed electron motion. Thus, RR effects can
enhance the ion acceleration in the optical transparency region \cite{Chen_2010_Radiation}, or
reduce it in dense plasma targets \cite{Capdessus_2015_Influence}.
The reduction of the ion peak energy and
of the width of the energy spectrum for linearly polarized laser pulses are shown in \cite{Tamburini_2010_Radiation}. Thus,
strong SEs can also be manifested  in the ion dynamics during laser-plasma interaction.

As known, the quantum radiation regime is determined by the invariant strong field quantum parameter
$\chi\sim 1$, and the SEs are distinct
in the quantum radiation-dominated regime (QRDR),  which requires $R_Q\equiv \alpha a_0\chi\gtrsim 1$ \cite{Piazza_2012_Extremely, Li_2014_Robust},
Here, $\chi = e\hbar \sqrt{(F_{\mu \nu}p^\nu)^2} / m^3c^4$
is the quantum invariant field parameter,
$\hbar$ the reduced Planck constant, $F_{\mu \nu}$  the electromagnetic field tensor, $p^\nu = (\varepsilon/c, \mathbf{p})$  the four-momentum of electron, $c$ the speed of light in vacuum,
 $\alpha = e^2 / c\hbar $ the fine structure constant, $a_0 = eE_0 / mc\omega_0$ the classical invariant  field parameter,
$\omega_0$ the laser frequency, $E_0$ the laser field amplitude, and $e$ and $m$ are the electron charge and mass, respectively.

\begin{figure}
 	\includegraphics[width=1.0\linewidth]{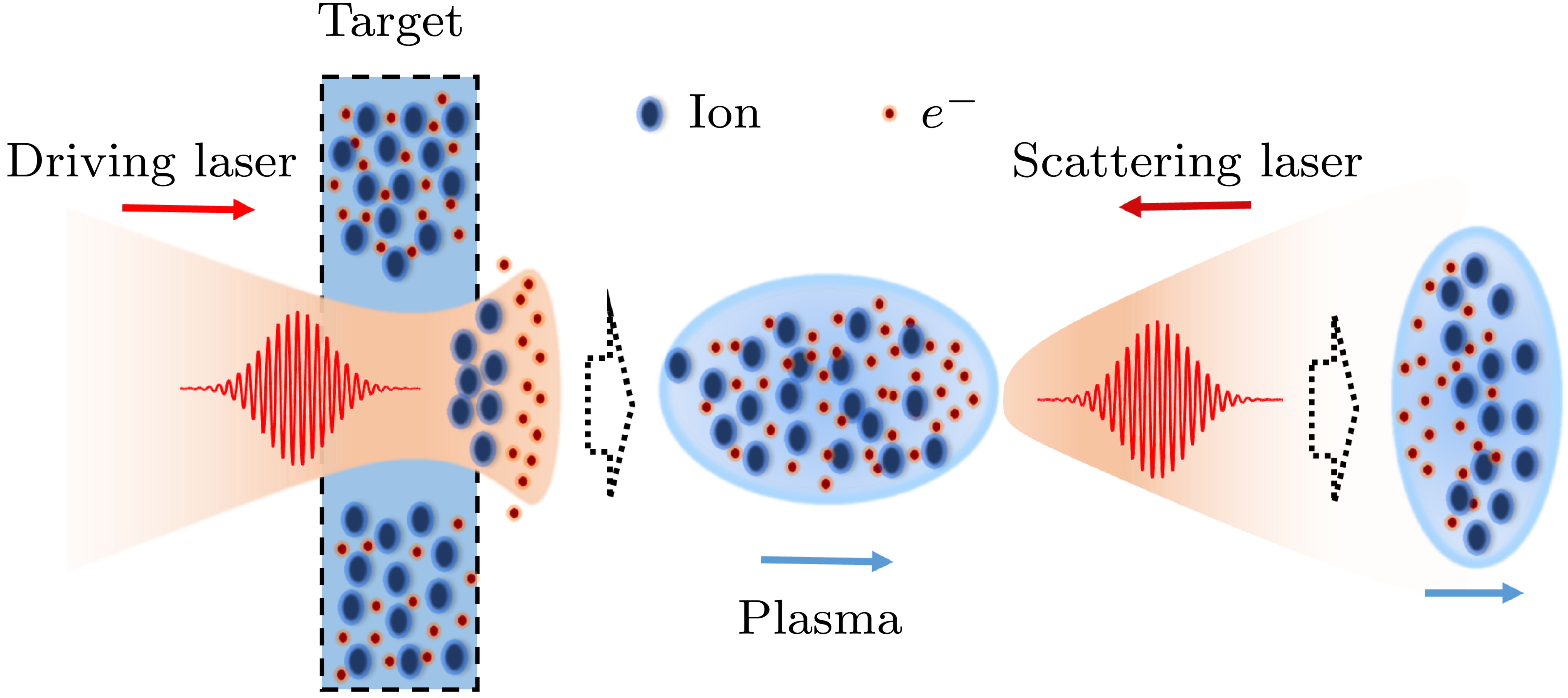}
\centering
	\caption{Scenario of detection of the SEs signature in the proton spectra. An ultrarelativistic plasma beam consisting of electrons and ions is generated by RPA, which further head-on collides with an ultraintense scattering laser pulse. The scattering laser pulse decelerates and compresses the electrons and, consequently, the protons because of the charge separation force.   As a result the proton energy spectrum acquires a SEs dependent downshift.
}\label{fig1}
\end{figure}

In this paper, we investigate theoretically the signature of SEs in the proton spectra during the interaction of a solid plastic target with laser fields in QRDR.
We consider a typical radiation pressure acceleration (RPA)  scheme, see \cite{Esirkepov_2004_Highly, Macchi_2005_Laser, Robinson_2008, Yu_2010, Steinke_2013,Aurand_2013,Kim_2016} and the references therein, when a circularly-polarized ultraintense laser pulse irradiates
a tens of nanometer solid target (e.g., the widely used polystyrene (PS)  $(\mathrm{C_8H_8})_n$ with an electron density about $10^3$ times higher than the critical density \cite{Limpouch_2004_Laser, Bugrov_2005_Experimental, Hilz_2018_Isolated}), and electrons are directly pushed out as a sheet by the light pressure \cite{Macchi_2005_Laser, Yan_2008_Generating, Qiao_2010_Radiation, Bulanov_2010_Unlimited}.   The charge separation field between the electron and ion sheets can accelerate ions continuously until the balance between this field and the radiation pressure breaks down or the plasma instabilities strongly deform the electron sheet \cite{Qiao_2010_Radiation, Bulanov_2010_Unlimited}.
  To detect SEs, the accelerated plasma
 head-on collides with another ultraintense (scattering) laser pulse,
 see the scenario in Fig.~\ref{fig1}. The electron sheet is slowed down  mostly due to RR, and partially pushed back by the scattering laser pulse, resulting in disturbance 
 of the acceleration progress of protons and compression of their energy spectra.

 We carry out optimization over the laser intensity and the pulse duration to achieve the best SEs signature. First of all, in the scattering stage the condition for the QRDR regime should be fulfilled $R_Q\gtrsim 1$ and $\chi \sim 1$. In the laser-electron counterpropagating configuration, $\chi$ can be estimated as $\chi \simeq 2\hbar \omega_0 \gamma_e a_0 / (mc^2)$, where  $\gamma_e$ is the electron Lorentz-factor.  Employing current achievable  laser pulses (the peak intensity $I_0\sim 10^{22} \mathrm{W/cm^2}$ and $a_0 \sim 10^2$) head-on colliding with an electron beam with a mean energy from hundreds of MeV to several GeV  ($\gamma_e \sim 10^2 - 10^3$) \cite{Esirkepov_2004_Highly, Pegoraro_2007_Photon,Bulanov_2010_Unlimited, Wang2013,Leemans2014,Wolter2016,Chatelain2014}, the QRDR condition $R_Q \gtrsim 1$ can be reached.
In the optimal case the peak intensity of the driving laser pulse is $8.5\times10^{22} \mathrm{W/cm^2}$ and the length is 6 cycles, while the scattering pulse peak intensity is $1.1\times10^{23} \mathrm{W/cm}^2$ and the length is 16 cycles. The mean energies of electrons and protons are both about hundreds of MeV, and the latter is
slightly higher.

We simulate the plasma dynamics through two semi-classical models based on the two-dimensional (2D) particle-in-cell (PIC) EPOCH code \cite{Ridgers_2012, Arber_2015}: the first considers quantum RR effects via the, so-called, modified Landau-Lifshitz (LL) equation \cite{Poder_2018_Experimental, Piazza_2012_Extremely}, which includes quantum recoil effects but no SEs; the second is a Monte-Carlo (MC) method, which includes both of quantum recoil effects and SEs \cite{Elkina_2011_QED, Ridgers_2014_Modelling, Green_2015_SIMLA, Gonoskov_2015_Extended}.
For the given parameters our simulations show that the SEs change significantly the shift of the proton energy spectra and the width of spectral distribution after scattering,
which can serve as a signature of SEs. The  compression of the proton energy distribution, seen previously beyond the QRDR in two-color \cite{Wan_2017_Stable} or two-intensity \cite{Zhou_2018_High} laser setups, is shown to be significant also in QRDR.

This paper is organized as follows. In Sec.~\ref{Sec2}, we present
theoretical models to describe the electron dynamics including RR effects. In Sec.~\ref{Sec3}, the signature of SEs on the proton energy spectra are presented and analyzed. In Sec.~\ref{Sec4}, the impact of laser and target parameters on the SEs signature are discussed. Finally, the conclusion is given in Sec.~\ref{Sec5}.

\section{Theoretical models}\label{Sec2}

 In the classical regime, i.e., $\chi\ll 1$, the LL equation \cite{Landau1975} can be employed to describe the electron dynamics including the RR effects. 
In the quantum regime $\chi \gtrsim 1$, this model 
overestimates the total radiation power, which is commonly handled by introducing a $g(\chi)$ function to suppress the RR force,  mimicking the effect of the quantum recoil  \cite{Kirk_2009_Pair, Sokolov_2010_Emission} (a polynomial fraction fit for $g(\chi)$ is given in Ref. \cite{Kirk_2009_Pair}). The modified LL (MLL) model includes quantum recoil effects but no SEs. However, at $\chi \gtrsim 1$, 
because of the discrete and probabilistic 
character of photon emission, the SEs can notably affect the electron dynamics and radiation \cite{Neitz_2014_Electron, Tamburini_2014_Electron, Bashinov_2015_Impact, Li_2017_Angle, Li_2018_Electron}, which is investigated in our MC simulations.

In ultraintense laser fields,  $a_0 \gg 1$, the coherence length of photon emission is much smaller than the laser wavelength and the typical size of the electron trajectory \cite{Ritus_1985_Quantum, Khokonov_2010_Length}. In this regime, 
the electron dynamics and radiation can be implemented via the MC simulation \cite{Elkina_2011_QED, Ridgers_2014_Modelling, Green_2015_SIMLA, Gonoskov_2015_Extended}.

\subsection{Modified Landau-Lifshitz (MLL) Model}

In the classical regime, the RR effects are described by the RR force inducing radiation damping. In contrast to the Lorentz-Abraham-Dirac (LAD) equation, the RR force in the LL equation \cite{Landau1975} is fully determined by the external fields
and avoids the runaway solutions \cite{Rohrlich_2008_Dynamics, Sokolov_2009_Dynamics, Hammond_2010_Radiation, Bulanov_2011}. The LL equation for the electron dynamics reads:
\begin{equation}
m \frac{du^{\mu}}{d\tilde{\tau}}=eF^{\mu \alpha}u_\alpha + f^{\mu},
\end{equation}
where the RR force is
\begin{eqnarray}\label{force}
f^{\mu}&=&\frac{2e^4}{3m^2c^4}(-F^{\mu \nu}F_{\nu \alpha}u^\alpha+(F^{\nu \beta}u_\beta F_{\nu \alpha}u^\alpha)u^\mu)\nonumber\\
&&+\frac{2e^3}{3mc^2}(\partial_\alpha F^{\mu \nu}u_\nu u^\alpha),
\end{eqnarray}
$u=(\gamma_e,\gamma_e \textbf{v}/c)$ is four-velocity of the electron, $\tilde{\tau}$ the proper time,
\begin{equation}
\frac{d}{d\tilde{\tau}}=c(k\cdot u)\frac{d}{d\tilde{\eta}},
\end{equation}
$\tilde{\eta}=(k\cdot \tilde{r})$,
and $\tilde{r}$ is the four-vector of the coordinate. The derivation of Eq.~(\ref{force}) is given in Chapt.~76 of Ref.~\cite{Landau1975}.
The three-dimensional expression of Eq.~(\ref{force}) reads \cite{Landau1975}:
\begin{widetext}
	\begin{eqnarray}\label{LL}
	{\bm F_{RR, classical}}&=&\frac{2e^3}{3mc^3} \left\{\gamma_e\left[\left(\frac{\partial}{\partial t}+\frac{\textbf{p}}{\gamma_e m}\cdot\nabla\right){\textbf{E}}+\frac{{\textbf{p}}}{\gamma_e m c}\times\left(\frac{\partial}{\partial t}+\frac{{\textbf{p}}}{\gamma_e m}\cdot\nabla\right){\textbf{B}}\right]\right.\nonumber\\
	&& +\frac{e}{m c}\left[{\textbf{E}}\times{\textbf{B}}+\frac{1}{\gamma_e m c}{\textbf{B}}\times\left({\textbf{B}}\times{\textbf{p}}\right)+\frac{1}{\gamma_e m c}{\textbf{E}}\left[{\textbf{p}}\cdot{\textbf{E}}\right)\right]\nonumber\\
	&& \left. -\frac{e\gamma_e}{m^2 c^2}{\textbf{p}}\left[\left({\textbf{E}}+\frac{{\textbf{p}}}{\gamma_e m c}\times {\textbf{B}}\right)^2-\frac{1}{\gamma_e^2m^2c^2}\left({\textbf{E}}\cdot{\textbf{p}}\right)^2\right]\right\},
	\end{eqnarray}
\end{widetext}
where $\textbf{E}$ and $\textbf{B}$ are the electric and magnetic fields, respectively.  Note that the last term in Eq.~(\ref{LL}) (proportional to $\gamma_e^2$) dominates over the preceding terms. From a practical point of view, the smaller terms may often be neglected, see Ref.~\cite{Tamburini_2010_Radiation}.

As $\chi\gtrsim 1$, the classical RR force in Eq.~(\ref{force}) overestimates the total radiation power, because quantum corrections decrease the average energy emitted by the electron compared to the classical one \cite{Ritus_1985_Quantum}. Nevertheless, the classical RR force can be modified in such a way that it will yield radiation losses corresponding to the quantum regime \cite{Kirk_2009_Pair, Sokolov_2010_Emission}. For this purpose the classical RR force is scaled by a function $g(\chi)$ to be suppressed appropriately at large $\chi$ \cite{Sokolov_2009_Dynamics,Thomas_2012_Strong}. The correspondingly modified RR force in the LL equation reads:
\begin{equation}
{\bm F_{RR, quantum}}=g(\chi){\bm F_{RR, classical}},\label{g_factor}
\end{equation}
where
\begin{eqnarray}
	g(\chi) 	&=& \frac{I_{QED}}{I_{C}}, \\
I_{QED}&=&mc^2\int c\left(k\cdot k' \right)\frac{d^2 W_{fi}}{d\tilde{\eta} dr_0}dr_0,\\
I_{C}&=&\frac{2e^4E'^2}{3m^2c^3},
\end{eqnarray}
$W_{fi}$ is the radiation probability, see below Eq.~(\ref{W}), $r_0=\frac{2\left(k\cdot k'\right)}{3\chi\left(k\cdot p_i\right)}$, and $E'$ is the electric fields in the electron rest frame. $k$, $k'$ and $p_i$ are the four-vectors of the wave vector of the driving laser, the wave vector of the radiated photon, and the momentum of the electron before the radiation, respectively. In the MLL equation, the recoil effects are included by rescaling the RR force by the factor $I_{QED}/I_{C}$, the ratio of the radiation intensities within QED and classical approaches, which will account for the classical overestimation of the RR effects for the electron dynamics.

\subsection{Monte-Carlo (MC) Model}

In this model, the calculation of the electron dynamics is based on the MC simulation employing QED theory for the electron radiation and classical equations of motion for the propagation of electrons between photon emission  \cite{Elkina_2011_QED, Ridgers_2014_Modelling, Green_2015_SIMLA, Gonoskov_2015_Extended}.
In superstrong laser fields, $a_0\gg 1$, the photon emission
probability $W_{fi}$ is determined by the local electron trajectory, consequently, by the local value of the parameter $\chi$ \cite{Ritus_1985_Quantum}:
\begin{eqnarray}
\frac{d^2 W_{fi}}{d \tilde{\eta} dr_0}=\frac{\sqrt{3}\alpha \chi [\int_{r_{\chi}}^{\infty} K_{5/3}(x)dx+ N_c K_{2/3}(r_{\chi})]}{2\pi\lambdabar_c(k\cdot p_i)},
\label{W}
\end{eqnarray}
where the coefficient $N_c=9r_0 r_{\chi} \chi^2/4$, the Compton wavelength $\lambdabar_c=\hbar/mc$, and $r_{\chi}  =r_0/(1-3\chi r_0/2)$.
The photon emission of electrons is considered to be a MC stochastic process \cite{Elkina_2011_QED, Ridgers_2014_Modelling, Green_2015_SIMLA, Gonoskov_2015_Extended}. During the electron-laser interaction, for each propagation coherent length $\Delta \tilde{\eta}$, the photon emission will take place if the condition $(dW_{fi}$/d$\tilde{\eta})\Delta\tilde{\eta}\geq N_{r}$ is fulfilled, where $N_r$ is a uniformly distributed random number in $[0, 1]$. Herein, the coherent length $\Delta \tilde{\eta}$ is inversely proportional to the invariant laser field parameter $a_0$, i.e., $\Delta \tilde{\eta}\sim 1/a_0$. However, to keep the total photon emission energy consistent, i.e., to exclude numerical error of the simulation of photon emission, we choose $\Delta \tilde{\eta} \ll 1/a_0$. The photon emission probability
\begin{eqnarray}
W_{fi} = \Delta \tilde{\eta}\frac{dW_{fi}}{d\tilde{\eta}}=\Delta \tilde{\eta}\int_{ \omega_{min}}^{ \omega_{max}} \frac{d^2 W_{fi}}{d \tilde{\eta} d\omega}d\omega, \nonumber
\end{eqnarray}
where $\hbar\omega_{min}$ and $\hbar\omega_{max}$ are assumed to equal the driving laser photon energy and  the electron instantaneously kinetic energy, respectively.
In addition, the emitted photon frequency $\omega_R$ is determined by the relation:
\begin{eqnarray}
\frac{1}{W_{fi}}\int_{\omega_{min}}^{\omega_R}\frac{d W_{fi}(\omega)}{d\omega}d\omega = \frac{\Delta\tilde{\eta}}{W_{fi}}\int_{\omega_{min}}^{\omega_R}\frac{d^2 W_{fi}(\omega)}{d\tilde{\eta} d\omega}d\omega=\tilde{N}_r,\nonumber
\end{eqnarray}
where, $\tilde{N}_r$ is another independent uniformly distributed random number in $[0, 1]$.

Between the photon emissions, the electron dynamics in the laser field is governed by classical equations of motion:
\begin{eqnarray}
\frac{d\bf{p}}{dt}=e({\bf E}+\frac{\textbf{v}}{c}\times\textbf{B}).
\end{eqnarray}
Given the smallness of the emission angle $ \sim 1/\gamma_e$  for an ultrarelativistic electron, the photon emission is assumed to be along the electron velocity. The photon emission induces the electron momentum change ${\bf p}_f \approx (1-\hbar\omega_R/c|{\bf p}_i|) {\bf p}_i$, where ${\bf p}_{i,f}$ are the electron momentum before and after the emission, respectively.

\section{Results and Analysis}\label{Sec3}

In the RPA scheme, multi-species plasmas with high electron density is beneficial for acceleration \cite{Yu_2010, Kar_2012, Kim_2013}. For the feasibility of experiments, the polystyrene (PS) $(\mathrm{C_8H_8})_n$ is chosen as the target, see Fig.~\ref{fig1}. This material can be easily manufactured in a relatively large range of density and has been widely used for laser-plasma interaction \cite{Limpouch_2004_Laser, Bugrov_2005_Experimental, Hilz_2018_Isolated}.
The target is an uniform plasma slab 
with  an electron density  $n_e = 300 n_c$  and thickness $l = 0.3\lambda_0$, where $n_c = m\omega_0^2 / 4 \pi e^2 $ is the critical density, and $\lambda_0=1 \mu$m the laser wavelength. 

In the 2D PIC simulations, we consider a simulation box of $60 \lambda_0\times 20 \lambda_0$ (in $z$ direction from $-20 \lambda_0$ to $40 \lambda_0$  and in $x$ direction from  $-10 \lambda_0$ to $10 \lambda_0$), and the corresponding cells are $3000\times 1000$.
Each cell is filled with 200 pseudo-electrons, 200 pseudo-protons and 100 $\mathrm{C^{6+}}$ macro-particles. The absorbing boundary conditions for the electromagnetic fields and particles  have been employed in transverse and longitudinal boundaries.
The driving laser pulse, propagating in the $+z$ direction, is circularly polarized with a profile at the left boundary ($z=-20\lambda_0$) of
\begin{eqnarray}
a_{driv}(x,t)&=&\frac{a_0}{2\sqrt{2}}\exp\left(-\frac{x}{w_0}^4\right)\left[\tanh\left[2\left(\frac{t}{T_0}-t_{01}\right)\right] \right.\nonumber\\
&&\left.-\tanh\left[2\left(\frac{t}{T_0} - t_{02}\right)\right]\right],\label{a_d}
\end{eqnarray}
where, $t_{01}=46$, $t_{02}=56$, the laser peak intensity $I_0 \approx 8.5\times10^{22} \mathrm{W/cm^2}$ (the total $a_0\approx250$), beam radius $w_0 = 5\lambda_0$, and $T_0$ is the laser period. Note that the distance from the left boundary of the simulation box to the solid target is 1 $\mu$m. The scattering laser pulse, propagating in the $-z$ direction,  is linearly polarized in the $x$ direction with a profile at the right boundary ($z=40\lambda_0$) of 
\begin{eqnarray}
a_{scat}(x,t)=a_1\exp\left[-\frac{x^2}{w_1^2}-\left(\frac{t}{T_0}-t_{11}\right)^2/t_{12}^2\right],\label{a_s}
\end{eqnarray}
where, $t_{11}=20$, $t_{12}=10$, the laser peak intensity  $I_1 \approx 1.1\times10^{23} \mathrm{W/cm}^2(a_1\approx286)$, wavelength $\lambda_1 = \lambda_0$, and beam radius $w_1 = 5\lambda_0$.

\begin{figure}[t]
	\includegraphics[width=1.0\linewidth]{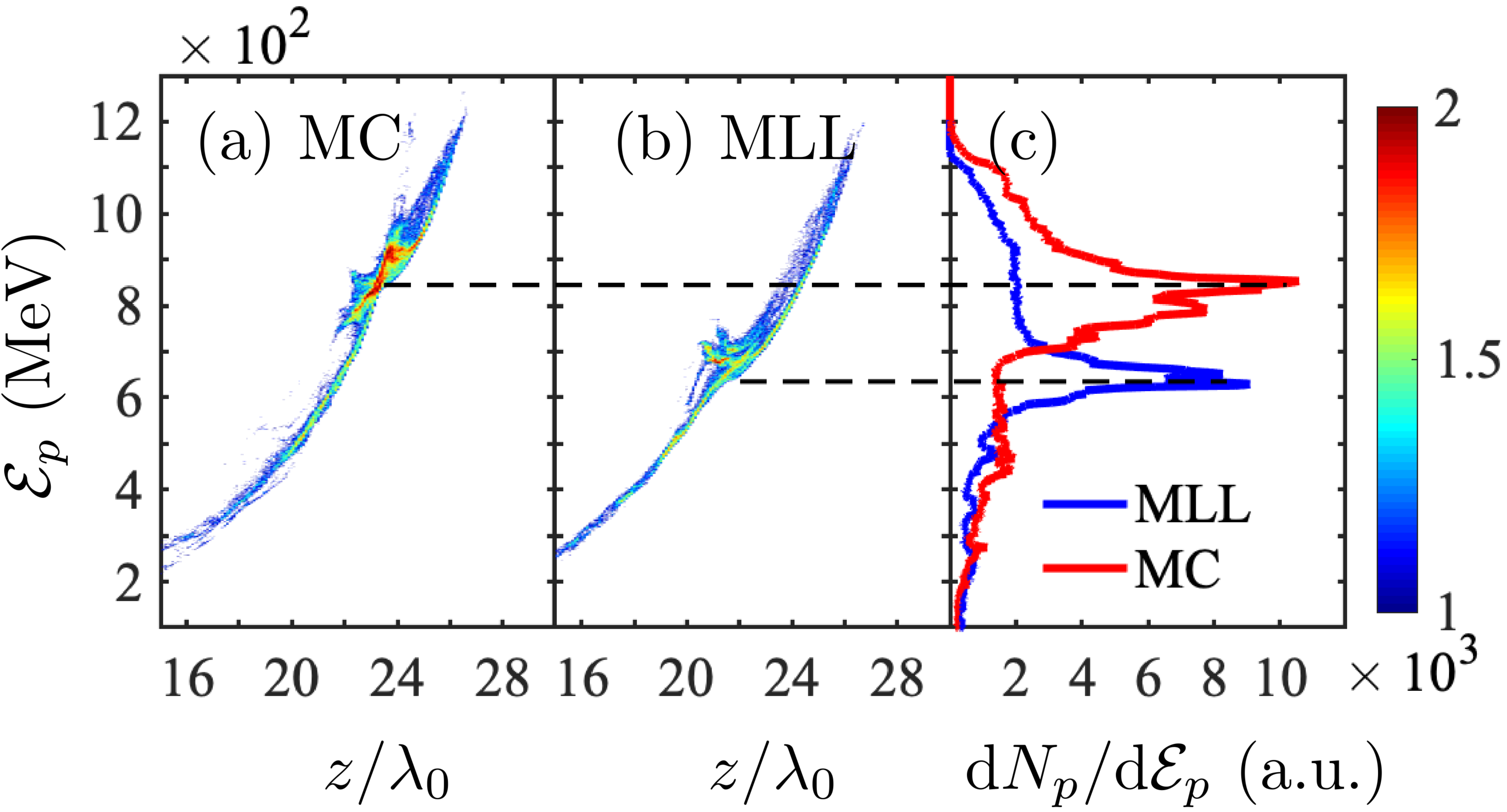}
\centering
	\caption{ (a) and (b): Proton energy spectra  Log$_{10}$[d$^2N_p/$d$\mathcal{E}_p$ d($z/\lambda_0$)] (a.u.) at the interaction time $t = 105 T_0$ vs the proton energy $\mathcal{E}_p$ and $z$ for  the MC and MLL models, respectively, where $N_p$ is the proton number;   (c):  d$N_p/$d$\mathcal{E}_p$ vs  $\mathcal{E}_p$. The red and blue curves indicate the MC and MLL models, respectively. 
The laser and plasma parameters are given in the text.} \label{fig2}
\end{figure}

A set of typical simulation results is presented in Fig.~\ref{fig2}. In the MC model including SEs, see Fig.~\ref{fig2}(a), a very dense proton layer is accelerated to the energy at the peak proton density of about 850 MeV in the longitudinal range of $23 \lambda_0 \lesssim z \lesssim 25 \lambda_0$. As the SEs are artificially removed in the MLL model, the energy at the peak proton density reaches only about 630 MeV, which is much lower than that in the  MC model, see Fig.~\ref{fig2} (b). The difference is more apparent in the $z$-integrated spectrum
shown in Fig.~\ref{fig2}(c). The energy shift on the proton energy spectra between the MLL and MC models can be considered as a signature of SEs because the difference between two models is that the
former excludes SEs, but the latter includes them.
In both cases, the acceleration process is disturbed by the scattering laser pulse.

\begin{figure}
    \includegraphics[width=\linewidth]{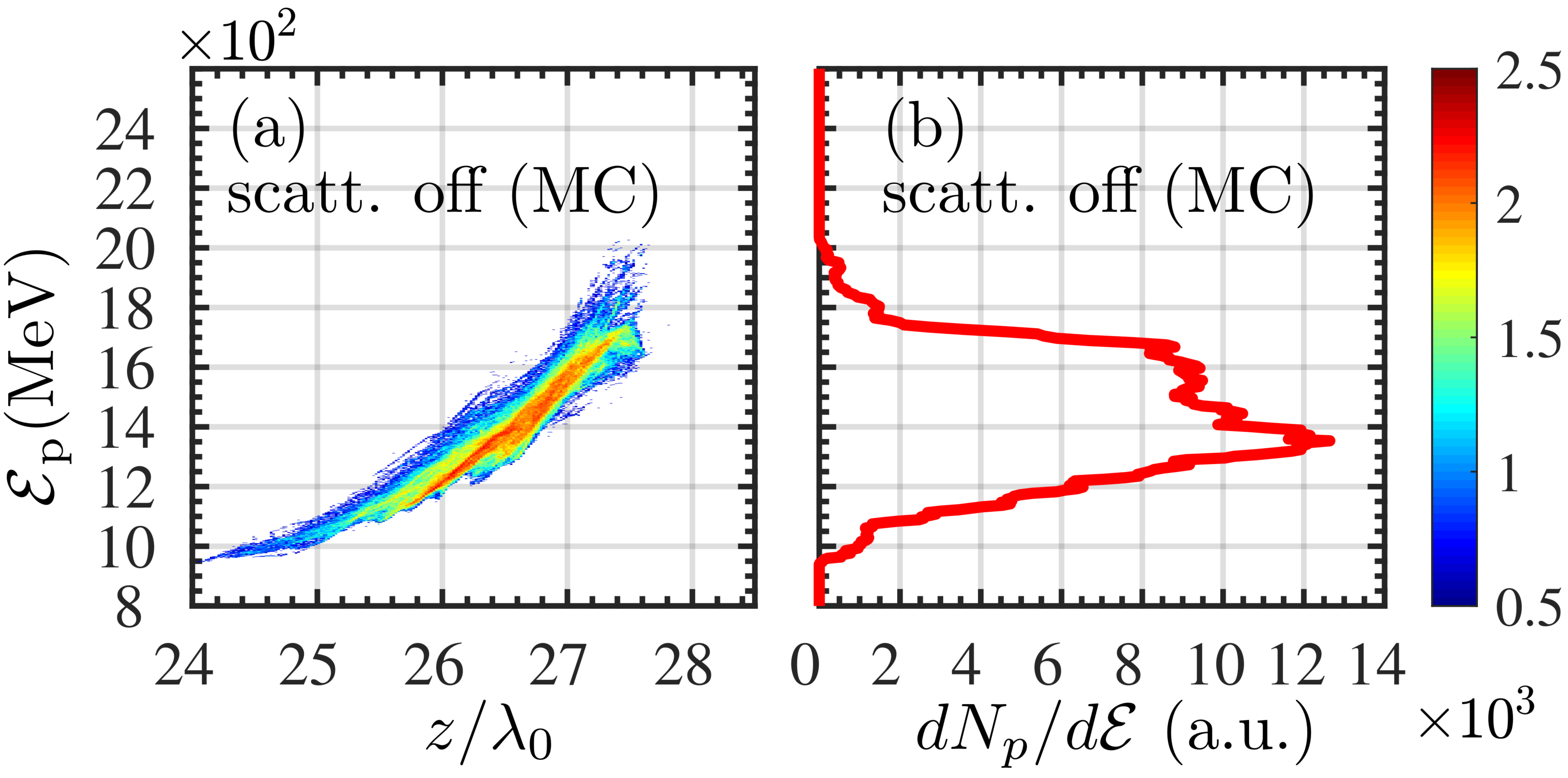}
    \caption{(a): Proton energy spectrum  Log$_{10}$[d$^2N_p/$d$\mathcal{E}_p$ d($z/\lambda_0$)] (a.u.) at $t = 105 T_0$ vs the proton energy $\mathcal{E}_p$ and $z$ for  the MC  model with scattering laser off. (b):  d$N_p/$d$\mathcal{E}_p$ vs  $\mathcal{E}_p$. The laser and target parameters are the same as those in Fig.~\ref{fig2}.} \label{fig3}
\end{figure}

\begin{figure}
    \includegraphics[width=\linewidth]{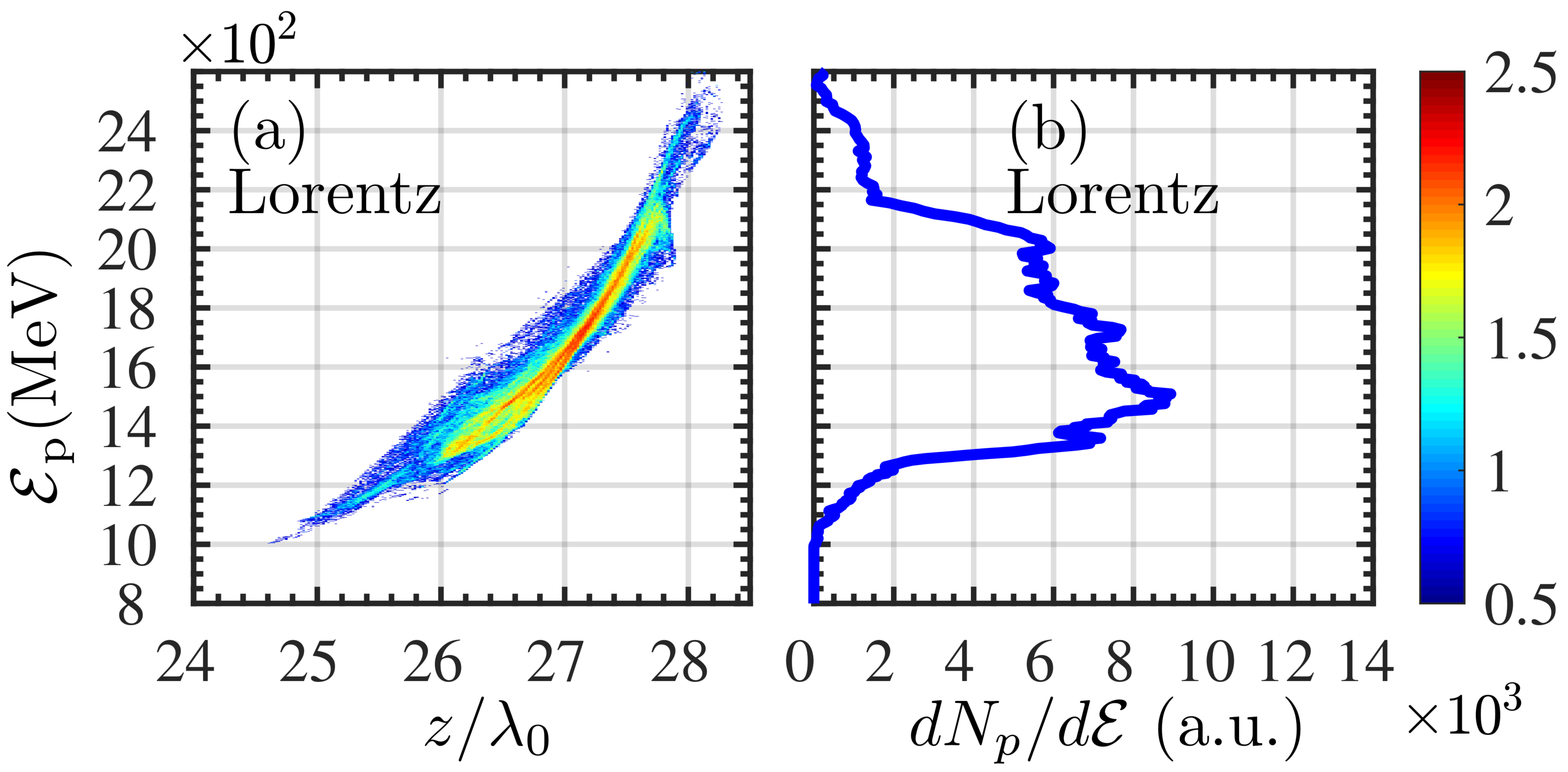}
    \caption{(a): Proton energy spectrum  Log$_{10}$[d$^2N_p/$d$\mathcal{E}_p$ d($z/\lambda_0$)] (a.u.) at $t = 105 T_0$ vs the proton energy $\mathcal{E}_p$ and $z$ for the Lorentz model excluding the RR effects. (b):  d$N_p/$d$\mathcal{E}_p$ vs  $\mathcal{E}_p$. The laser and target parameters are the same as those in Fig.~\ref{fig2}.} \label{fig4}
\end{figure}

\begin{figure*}
\center
\includegraphics[width=1\linewidth]{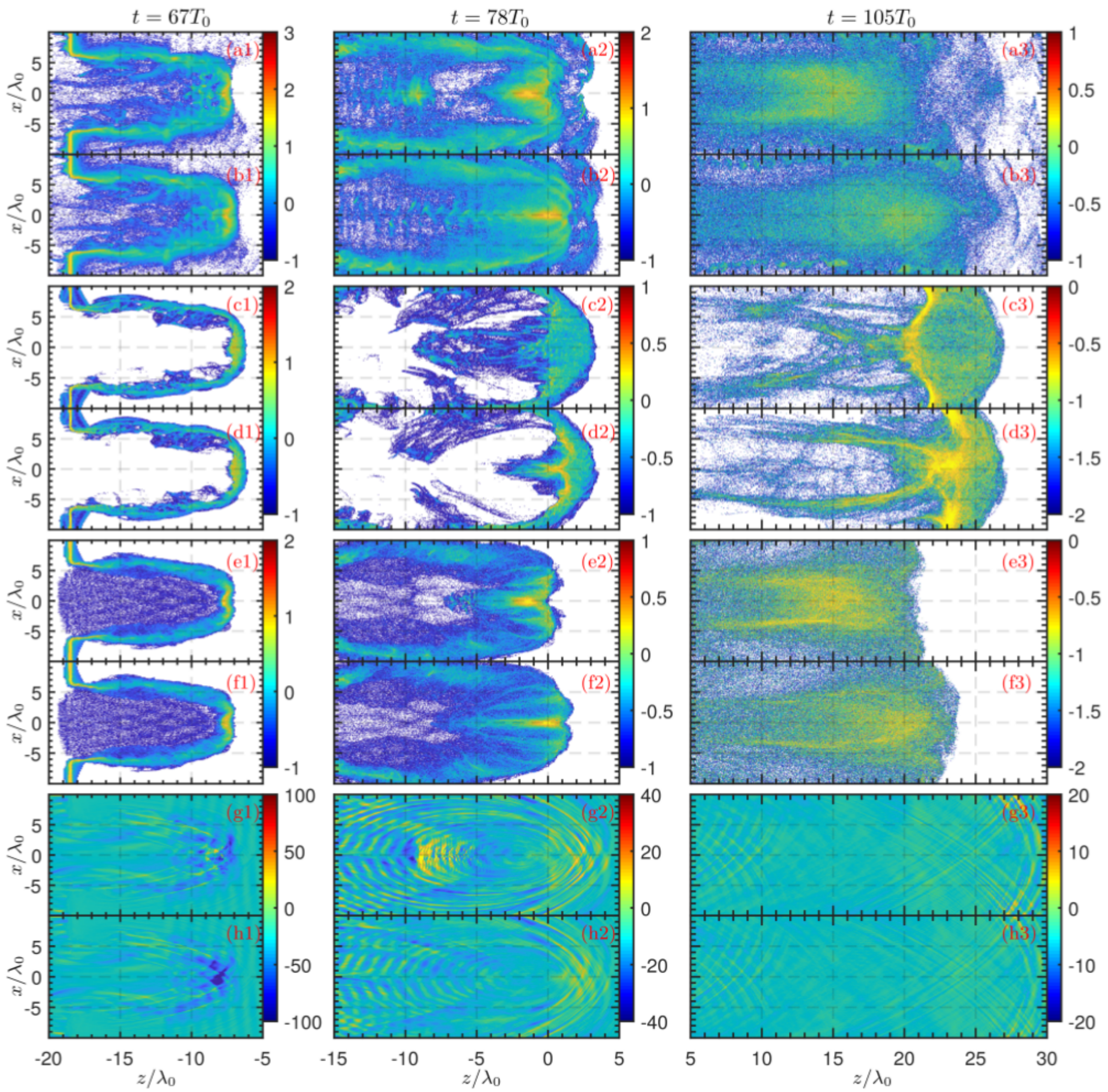}
	\caption{(a1)-(a3) and (b1)-(b3): Density distributions of electrons $\log_{10}(n_e/n_c)$ simulated via MLL and MC models, respectively. (c1)-(c3) and (d1)-(d3): Density distributions of protons $\log_{10}(n_p/n_c)$ simulated via MLL and MC models, respectively. (e1)-(e3) and (f1)-(f3): Density distributions of $C^{6+}$ $\log_{10}(n_C/n_c)$ simulated via MLL and MC models, respectively. (g1)-(g3) and (h1)-(h3): Distributions of the longitudinal electric field $E_z$  simulated via MLL and MC models, respectively. The interaction times in left, middle and right columns are $t=67 T_0$, $78 T_0$ and $105 T_0$, respectively.
		Other parameters are the same as those in Fig.~\ref{fig2}. }
		\label{fig5}
\end{figure*}

Figure~\ref{fig3} indicates the impact of the scattering laser pulse on the proton energy spectrum,
when compared with Fig.~\ref{fig2}. As the scattering laser pulse is not employed, the protons can be accelerated to a much higher energy: the energy at the peak density is about 1.35 GeV, and the energy spreading is much broader, see the detailed explanation in Fig.~\ref{fig5}.
Moreover, one can see from Fig.~\ref{fig4} that compared with Fig.~\ref{fig2} the RR effects also induce compression of the proton spectrum. 
In fact, without RR the proton spectrum is much broader than with RR.

\begin{figure}
    \includegraphics[width=\linewidth]{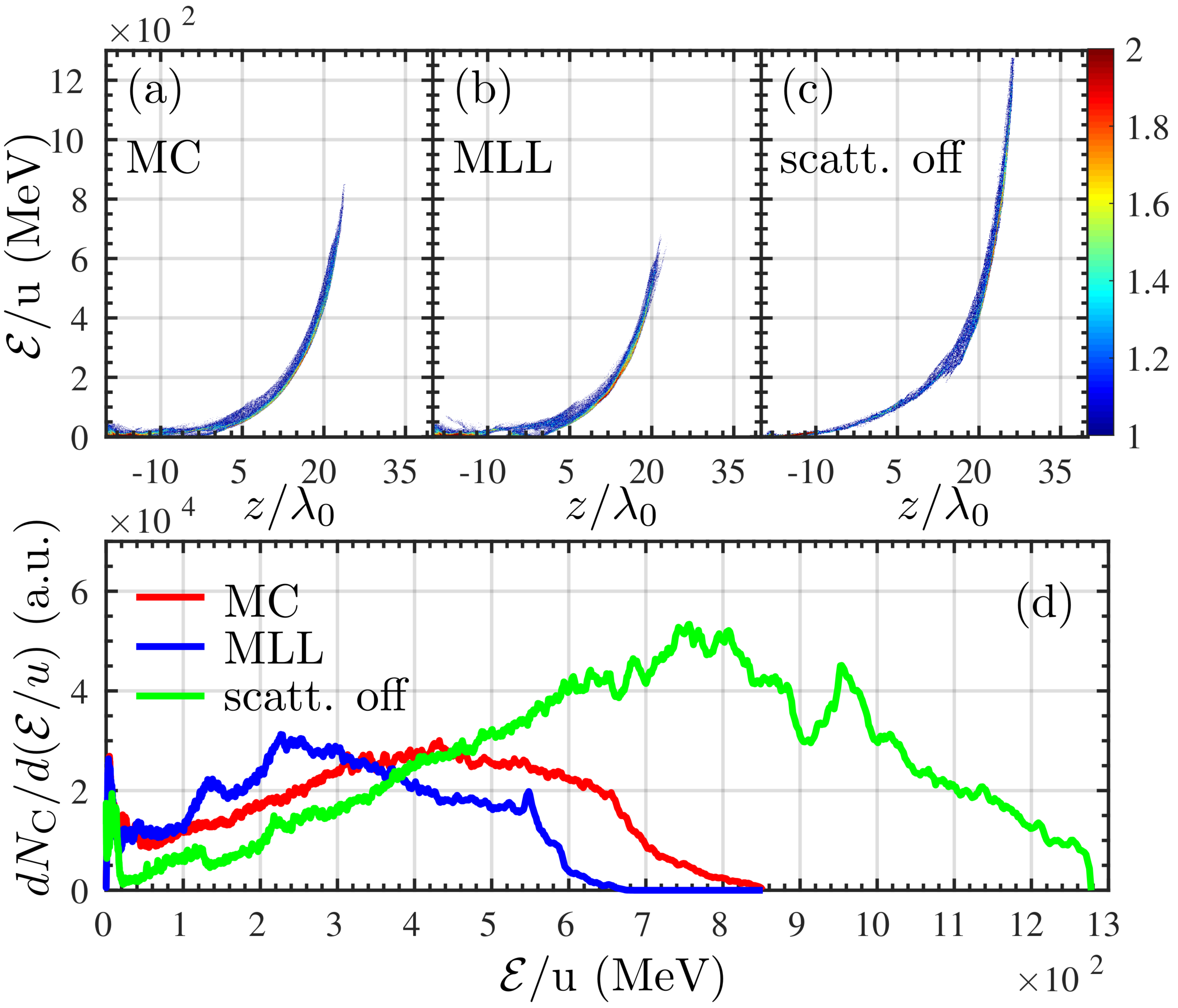}
    \caption{(a) and (b):  $\mathrm{C^{6+}}$ ion energy spectra Log$_{10}$[d$^2N_C/$ d$(\mathcal{E}/u)/$d($z/\lambda_0$)] (a.u.) at $t=105T_0$. with respect to $\mathcal{E}/u$ and $z$ in  MC and MLL models, respectively. The nucleon number $u= 12$ for $\mathrm{C^{6+}}$ ions. (c):  $\mathrm{C^{6+}}$ ion energy spectra Log$_{10}$[d$^2N_C/$d$(\mathcal{E}/u)/$d($z/\lambda_0$)] (a.u.) in the MC model without the scattering laser pulse. (d): $\mathrm{C^{6+}}$ ion energy spectra d$N_C/$d$(\mathcal{E}/u)$ with respect to $\mathcal{E}/u$. Other parameters are the same as those in Fig.~\ref{fig2}.}
    \label{fig6}
\end{figure}

 As already mentioned the scattering laser pulse can disturb the acceleration process, i.e., the balance between the electrostatic field and the light pressure. Comparing Figs.~\ref{fig2} and \ref{fig3},  we see that the scattering laser pulse induces decrease of the mean proton energy as well the energy bandwidth of the beam. In the  MLL model the downshift of the proton spectrum is much larger than in the MC model, which means a larger disturbance on the acceleration process.
The time-resolved dynamics of electrons, protons $\mathrm{C^{6+}}$ ions and longitudinal electric field in these models are analyzed in Fig.~\ref{fig5}. At the beginning of scattering, see Figs.~\ref{fig5}(a1) and (b1), slight differences on electron dynamics in the  MC and MLL models emerge already. The electron density in the region of $-8\lesssim z/\lambda_0 \lesssim -6$ in the MLL model (Fig.~\ref{fig5}(a1)) is lower than in the MC model (Fig.~\ref{fig5}(b1)) and excites different longitudinal force, see Figs.~\ref{fig5}(g1) and (h1). The corresponding proton density distributions are shown in Figs.~\ref{fig5}(c1) and (d1). For the given parameters in  the  MLL model more electrons can be scattered back than in the MC model 
with SEs \cite{Li_2017_Angle, Li_2018_Electron, Geng_2018_Quantum}. The electrons are reflected by the scattering laser pulse when the reflection condition  $\gamma_e\lesssim a_1/2$ \cite{Piazza_2012_Extremely} is fulfilled because of the radiation energy loss. While in the MLL model all electrons in a certain phase space have the same radiative loss, in the MC model the photon emission is a stochastic process and, as a result, the radiative losses for some electrons are suppressed and they penetrate further forward \cite{Geng_2018_Quantum}.
Thus, the  electron peak density  is higher in the MC model than in the MLL model, cf. Figs.~\ref{fig5}(a2) and (b2). More unscattered electrons result in changing the charge separation force, different longitudinal field and proton distributions, see Figs.~\ref{fig5}(c2), (d2), (g2) and (h2). 
After the scattering, those differences are further enhanced due to the plasma evolution, see  Figs.~\ref{fig5}(a3), (b3), (c3), (d3),  (g3) and (h3). Thus, the absence of SEs in the MLL model can induce a downshift of the proton energy spectrum compared with that in the MC model.

\begin{figure}[t]
	\includegraphics[width=\linewidth]{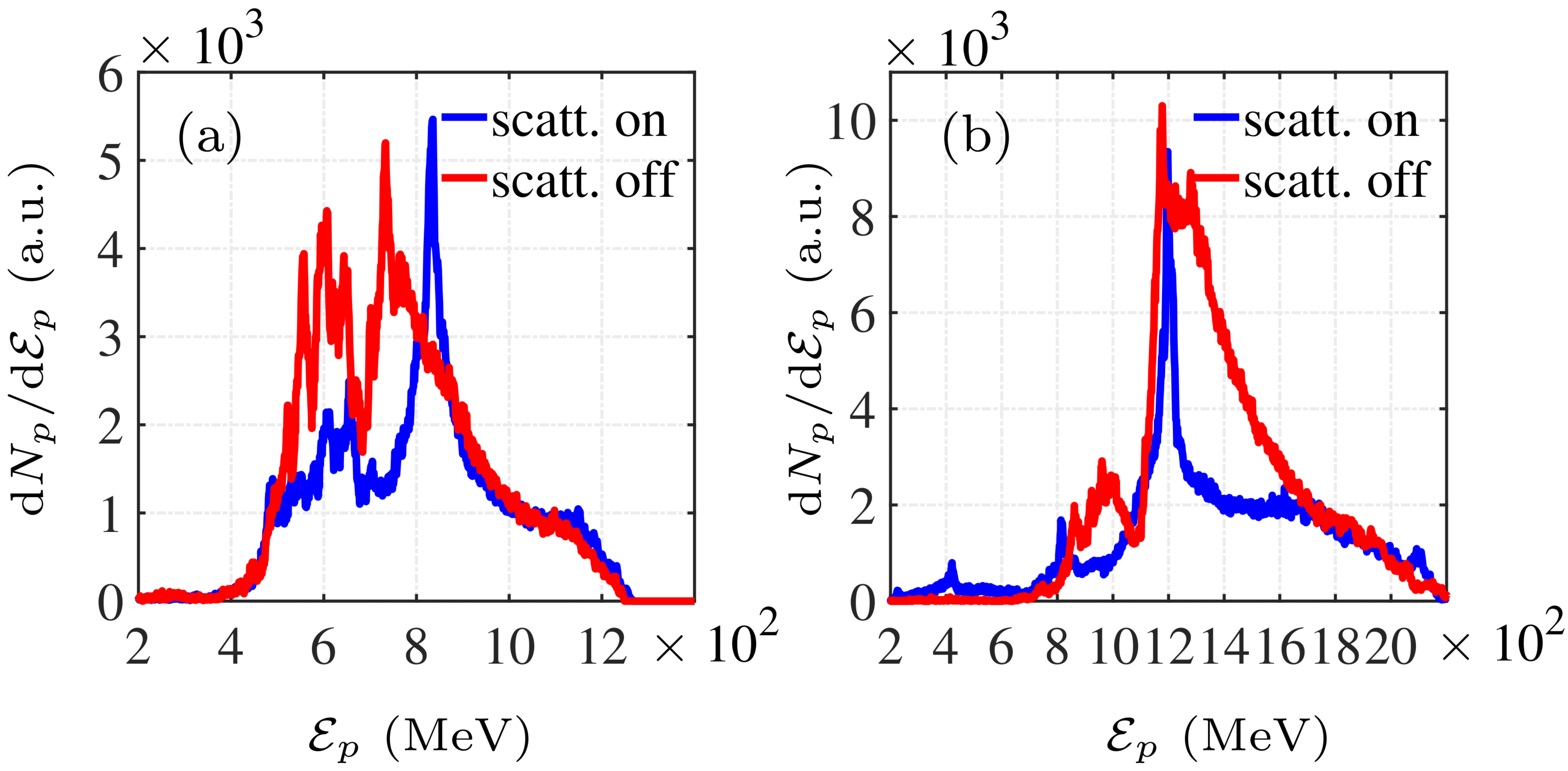}
	\centering
	\caption{Spectrum compression effect at $t\approx140.7T_0$ with $t_{01}=36$ and $t_{02}=43$ in Eq.~\ref{a_d}. (a) $a_0 \approx200$, $a_1 \approx 175$, and $n_e / n_c =  200$. (b) $a_0 \approx300$, $a_1 \approx 265$, and $n_e / n_c =  300$.  The blue and red curves indicate including and excluding the scattering laser pulse. Here, the simulation box is $120 \lambda_0\times 20 \lambda_0$ (in $z$ direction from $-20 \lambda_0$ to $100 \lambda_0$  and in $x$ direction from  $-10 \lambda_0$ to $10 \lambda_0$), and the corresponding cells are $6000\times 1000$. Other parameters are the same as those in Fig.~\ref{fig2}.}
	 \label{fig7}
\end{figure}

The dynamics of both protons and $\mathrm{C^{6+}}$ ions are mostly governed by electrons. For instance, as shown in Figs.~\ref{fig5}(e1)-(e3) and (f1)-(f3). At $t=78T_0$ and $105T_0$ the protons and $\mathrm{C^{6+}}$ ions have been split apparently because the latter are much heavier and slower. 
The energy spectrum of $\mathrm{C^{6+}}$ ions at $t=105T_0$, see Fig.~\ref{fig6}, show similar behaviours as in the case of protons in Figs.~\ref{fig2} and \ref{fig3}. The absence of SEs in the MLL model induces a downshift of the energy spectrum of $\mathrm{C^{6+}}$ ions, compared with that in the MC model, see Figs.~\ref{fig6}(a) and (b). The previous has a peak energy of about 226 MeV per nucleon with a FWHM of about 444 MeV, and the latter has a peak energy of about 432 MeV per nucleon with a FWHM of about 503 MeV, as shown by the blue and red curves in Fig.~\ref{fig6}(d). When the scattering laser pulse is switched off, the $\mathrm{C^{6+}}$ ions achieve a much higher energy, but with much broader energy spreading. For instance, as shown in Figs.~\ref{fig6}(c) and (d), the $\mathrm{C^{6+}}$ ions have a peak energy of about 756 MeV per nucleon with a FWHM of about 606 MeV.

Furthermore, we underline that our configuration can also be employed to compress the proton energy spectra, as shown in Fig.~\ref{fig7}, in the QRDR (beyond the QRDR, similar setups have been considered to compress the proton energy spectra \cite{Wan_2017_Stable,  Zhou_2018_High}). As the front electrons in the accelerated plasma are decelerated or even reflected by the scattering laser pulse due to the radiation energy loss, the front protons are decelerated  due to the charge separation force. Consequently, the proton energy spectra are significantly compressed. For instance, in Fig.~\ref{fig7}(a) the peak energies without and with the scattering laser pulse are $\mathcal{E}_p^{peak} \approx$ 732 MeV and 835 MeV, respectively, and the scattering laser pulse can significantly compress the energy spreading from  45.5\% to 7.9\%; in Fig.~\ref{fig7}(b), without and with the scattering laser pulse, the peak energies $\mathcal{E}_p^{peak} \approx$ 1177 MeV and 1197 MeV, respectively, and the corresponding energy spreadings are about 23.9\% and 4.1\%, which may be very useful for studies and applications of ion acceleration.

\section{The impact of laser and target parameters on the signature of SEs}\label{Sec4}

\begin{figure}[t]
	\includegraphics[width=1.0\linewidth]{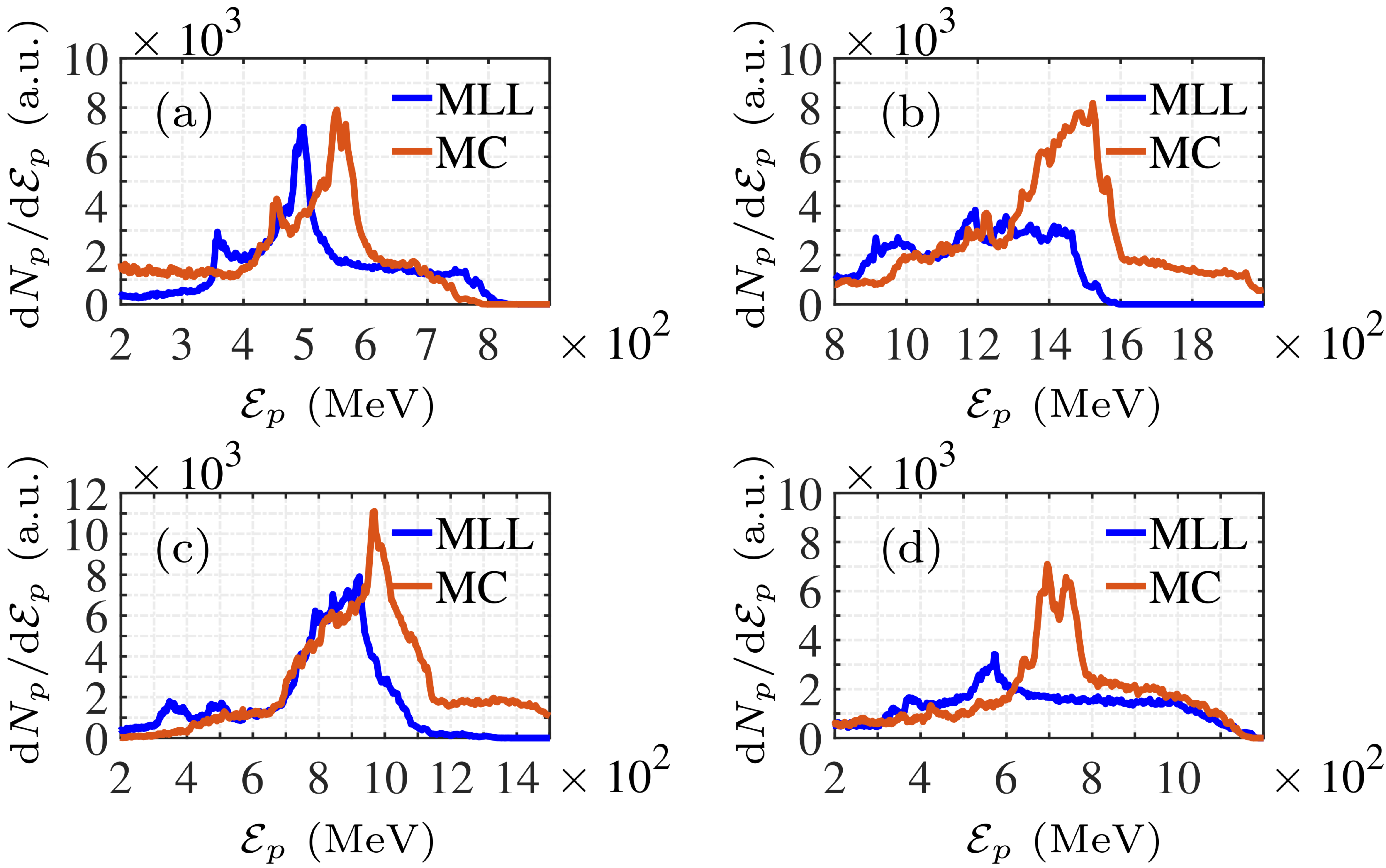}
	\centering
	\caption{Proton energy spectra at $t=105 T_0$. The driving laser peak intensity in (a) is $I_{0} \approx 5.4 \times 10^{22} \mathrm{W/cm}^2 (a_0\approx200)$ and (b) $I_{0} \approx 1.2 \times 10^{23}\mathrm{W/cm}^2 (a_0\approx300)$, respectively. The scattering laser peak intensity in (c) is $I_{1} \approx 7.7 \times 10^{22}\mathrm{W/cm}^2 (a_1\approx238)$ and (d) $I_{1} \approx 1.5 \times 10^{23}\mathrm{W/cm}^2 (a_1\approx332)$, respectively. Other parameters are the same as those in Fig.~\ref{fig2}. }\label{fig8}
\end{figure}

 In this section we analyze the optimal conditions in terms of the laser and the target parameters for future experimental studies.

\subsection{The role of the laser intensities}

We start with the intensities of the driving laser $a_0$ and the scattering laser $a_1$. As we know already, the signature of SEs is observed via the shift of the proton energy spectra, when $a_0$ and $a_1$ match each other within a certain range, see Fig.~\ref{fig8}. Decreasing the driving laser intensity $a_0$ will reduce the energy gain of electrons and protons in RPA, consequently, the SEs are suppressed due to the decrease of the parameter $\chi\sim\gamma_e  a_1$ during the interaction with the scattering laser pulse.  A sample of $a_0 \approx 200$ is shown in Fig.~\ref{fig8}(a). The peak energy $\mathcal{E}_p^{peak}$ shift between MC ($\mathcal{E}_p^{peak}\approx 552.4$ MeV with a FWHM of about 72.2 MeV) and MLL ($\mathcal{E}_p^{peak}\approx 497.7$ MeV with a FWHM of about 53 MeV) reduces to about 54.7 MeV, compared with that in Fig.~\ref{fig2} ($a_0 \approx 250$).

For a larger $a_0$, on one hand the SEs are stronger, because of the higher electron energy and the higher parameter $\chi$;  on another hand, due to a larger $a_0$ the driving laser pulse can still provide acceleration  when the scattering process is finished, which may overwhelm SEs, see Fig.~\ref{fig8}(b). In the MC model, $\mathcal{E}_p^{peak}\approx 1522$ MeV with a FWHM of about 250 MeV, and $\Delta \mathcal{E}_p/\mathcal{E}_p\approx16.43$\%, which is much less that of about 48.37\% in the MLL model ($\mathcal{E}_p^{peak}\approx 1193$ MeV with a FWHM of about 577 MeV).

A similar trade-off exists for the scattering laser intensity $a_1$.
If $a_1$ is weak, the parameter $\chi\sim\gamma_e a_1$ for electrons is small as well, and the SEs are suppressed, see the case for $a_1 \approx 250$ in Fig.~\ref{fig8}(c). In the MC model, $\mathcal{E}_p^{peak}\approx 968.2$ MeV with a FWHM of about 244 MeV; In the MLL model, $\mathcal{E}_p^{peak}\approx 923$ MeV with a FWHM of about 210 MeV. On the contrary, a stronger $a_1$ can enhance SEs, thus, more electrons can be scattered back, see the case of $a_1 = 350$ in Fig.~\ref{fig8}(d). In the MC model, $\mathcal{E}_p^{peak}\approx 695.7$ MeV with a FWHM of about 105 MeV; In the MLL model, $\mathcal{E}_p^{peak}\approx 574.5$ MeV with a FWHM of about 177 MeV. However, a much stronger $a_1$  not only enhances the SEs but also pushes all electrons backward, which may 
weaken the SEs signature.

\subsection{The role of the laser pulse durations}

\begin{figure}[t]
	\includegraphics[width=1.0\linewidth]{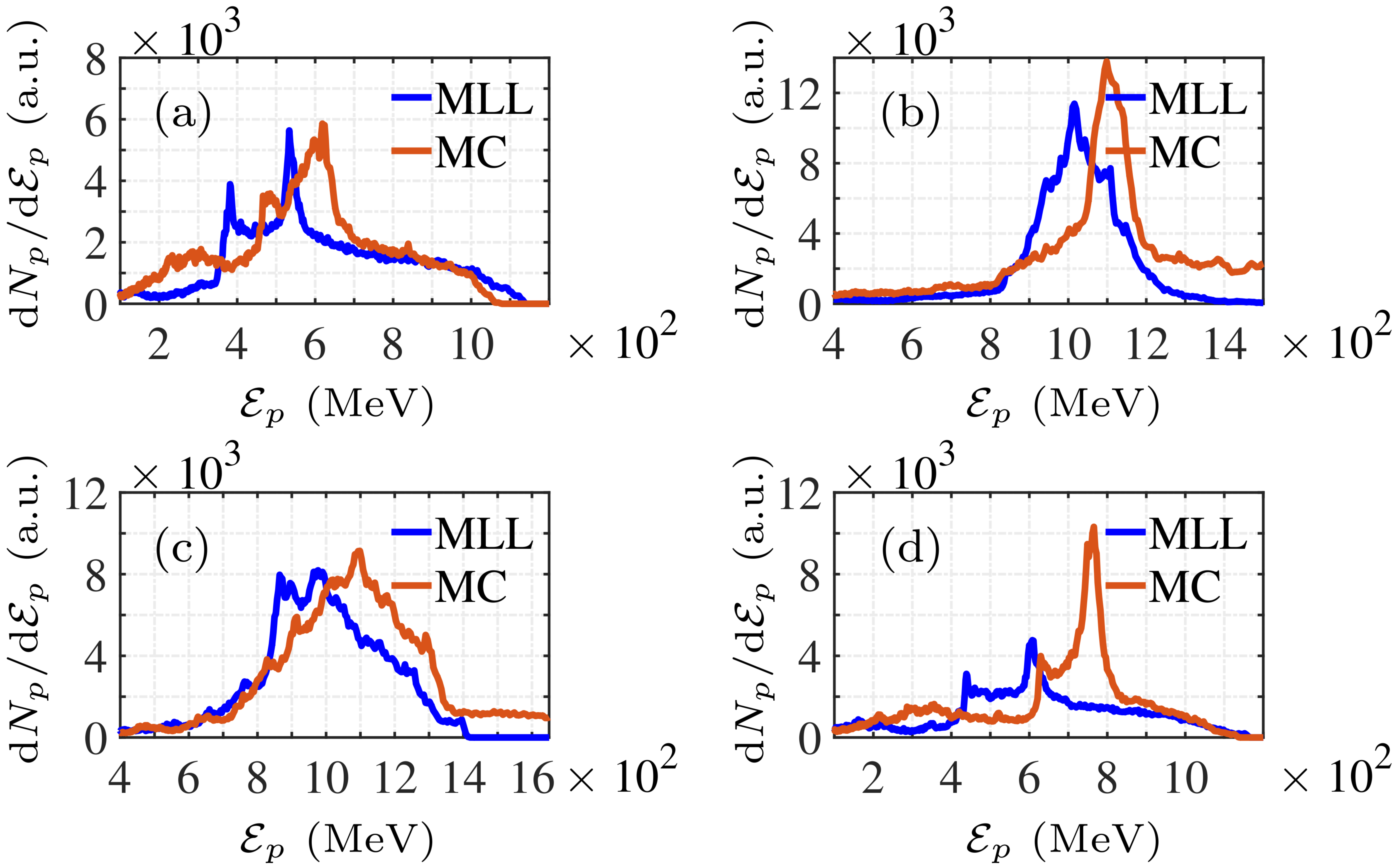}
	\centering
	\caption{Proton energy spectra at $t=105 T_0$. The pulse duration of the driving laser pulse in (a) is $7T_0$ with $t_{01} = 46$ and $t_{02}=53$, and in (b) $15 T_0$ with $t_{01} = 46$ and $t_{02}=61$. The pulse duration of the scattering laser in (c) is set with $t_{11} = 20$ and $t_{12}=5$, and in (d)  with $t_{11} = 20$ and $t_{12}=15$. See the definitions of $t_{01}$, $t_{02}$, $t_{11}$ and $t_{12}$ in Eqs.~(\ref{a_d}) and (\ref{a_s}). Other parameters are the same as those in Fig.~\ref{fig2}. }\label{fig9}
\end{figure}

The role of the pulse durations of the driving laser $\tau _0$ and the scattering laser $\tau _1$ are analyzed in Fig.~\ref{fig9}.  For a shorter $\tau_0$, the electron energy is lower because of shortness of the relevant acceleration time, and the SEs are more difficult to be distinguished since $\chi\sim\gamma_e a_1$, see Fig.~\ref{fig9}(a). The latter shows that the energy shift between MC and MLL is smaller compared with that in Fig.~\ref{fig2}. In the MC model, $\mathcal{E}_p^{peak}\approx 618.6$ MeV with a FWHM of about 134.8 MeV; In the MLL model, $\mathcal{E}_p^{peak}\approx 533.5$ MeV with a FWHM of about 52 MeV.

For a longer $\tau_0$, the electrons scattered due to the RR effects can be re-accelerated to high energies, yielding to the suppression of SEs, see the sample case  in Fig.~\ref{fig9}(b). In the MC model, $\mathcal{E}_p^{peak}\approx 1098$ MeV with a FWHM of about 104 MeV; In the MLL model, $\mathcal{E}_p^{peak}\approx 1016$ MeV with a FWHM of about 184 MeV. 

The duration of the scattering laser pulse determines the total energies of electrons depleted by the RR effects. 
Shortening of the pulse duration of the scattering laser can reduce the radiation and subsequent SEs, see the case  in Fig.~\ref{fig9}(c). In the MC model, $\mathcal{E}_p^{peak}\approx 1098$ MeV with a FWHM of about 374 MeV; In the MLL model, $\mathcal{E}_p^{peak}\approx 976.7$ MeV with a FWHM of about 328.3 MeV. A longer $\tau _1$ can  strengthen the RR effects and also SEs, see Fig.~\ref{fig9}(d). In the MC model, $\mathcal{E}_p^{peak}\approx 765.5$ MeV with a FWHM of about 60 MeV; In the MLL model, $\mathcal{E}_p^{peak}\approx 610.6$ MeV  with a FWHM of about 66 MeV.

\subsection{The role of the time delay between the laser pulses}

The effects of the time delay between the laser pulses 
are analyzed in Figs.~\ref{fig10}(a) and (b). If the time delay is increased, i.e., the driving laser pulse enters earlier,
the protons and electrons can be accelerated to higher energies. This leads to a stronger RR force since $\chi\sim\gamma_e a_1$. However, if the delay is long enough for completing the acceleration process, such as the case  in Fig.~\ref{fig10}(a), it will be hard for the scattering laser pulse of the given intensity to decelerate the electrons, to significantly disturb  
the proton spectra, and consequently, to resolve SEs.
  In the MC model, $\mathcal{E}_p^{peak}\approx 1016$ MeV with a FWHM of about 94.1 MeV; In the MLL model, $\mathcal{E}_p^{peak}\approx 772.2$ MeV with a FWHM of about 594 MeV.

If the time delay is shorter, i.e., the driving laser pulse enters later,
the acceleration process is disturbed earlier. This prolongs the interaction between the driving laser pulse and the scattered plasma. Consequently, the final SEs 
on the spectra 
is larger, see the case  in Fig.~\ref{fig10}(b). In the MC model, $\mathcal{E}_p^{peak}\approx 772.2$ MeV with a FWHM of about 292 MeV; In the MLL model, $\mathcal{E}_p^{peak}\approx 554.8$ MeV with a FWHM of about 117 MeV.

\subsection{The role of the target density}

\begin{figure}[t]
	\includegraphics[width=1.0\linewidth]{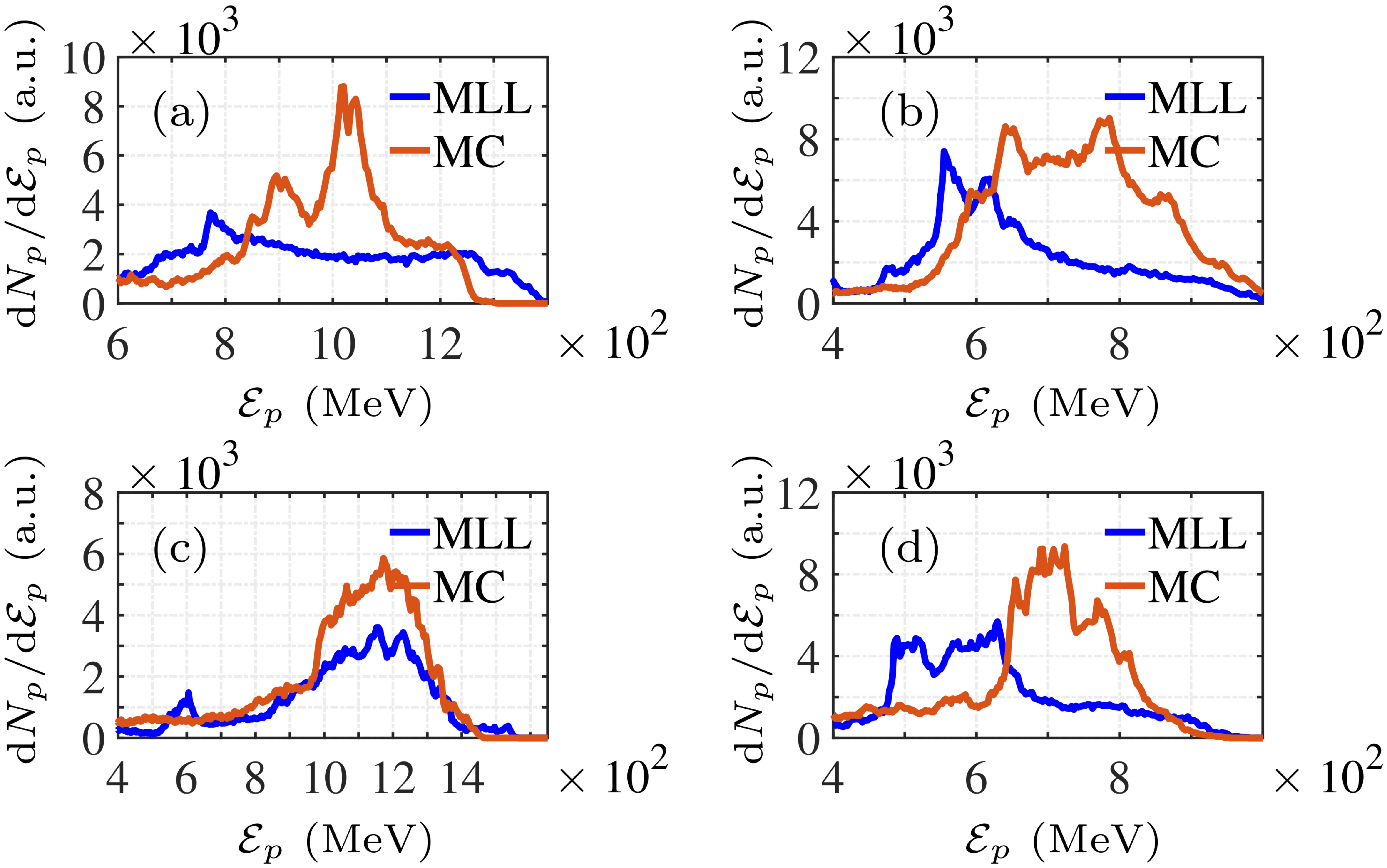}
	\centering
	\caption{Proton energy spectra at $t=105T_0$. The driving laser enters $10T_0$ earlier in (a) with $t_{01} = 36$ and $t_{02}=46$,  and $10T_0$ later in (b) with $t_{01} = 56$ and $t_{02}=66$. $t_{11}$ and $t_{12}$ keep constant.  (c) and (d): Initial density of electrons $n_e = 250n_c$ and $350n_c$, respectively. Other parameters are the same as those in Fig.~\ref{fig2}. }\label{fig10}
\end{figure}

The role of the target density $n_e$ is analyzed in Figs.~\ref{fig10}(c) and (d). The reduction of the electron density can reduce the acceleration efficiency and also the scattered electron number, leading to the SEs  suppression. For instance,  for the case of $n_e=250n_c$, the signature of SEs cannot be apparently observed, see Fig.~\ref{fig10}(c).  In the MC model, $\mathcal{E}_p^{peak}\approx 1172$ MeV with a FWHM of about 318 MeV; In the MLL model, $\mathcal{E}_p^{peak}\approx 1157$ MeV with a FWHM of about 333 MeV.

For a higher density target, such as the case of $n_e=350n_c$ in Fig.~\ref{fig10}(d), acceleration effects are lower due to the larger target mass and yield lower spectra \cite{Qiao_2010_Radiation}, however, the SEs can still be resolved.  In the MC model, $\mathcal{E}_p^{peak}\approx 723.4$ MeV with a FWHM of about 147 MeV; In the MLL model, $\mathcal{E}_p^{peak}\approx 629.4$ MeV with a FWHM of about 179 MeV. For much higher densities, much stronger laser pulses are required for acceleration, otherwise, collective RR effects on protons are rather weak.

Finally, note that in our simulations the transverse Rayleigh-Taylor (RT) like instability \cite{Ott1972, Pegoraro2007} exists in both of the MLL and MC models, and the difference between these two models is the stochasticity effects. Thus, the comparisons of the MLL and MC simulations distinguishes explicitly the role of SEs, but not the role of the RT-like instability. Our estimations show that in the rest frame of the moving RPA plasma the electrostatic field $E'_{RT}$ induced by the RT-like instability is much smaller than the scattering laser field $E'_1$:  $E_1'/E'_{RT}\sim \gamma_e^2a_1/(\frac{n_e}{n_c}\frac{l_{RT}}{\lambda_1})$, where $\lambda_1$ is the wavelength of the scattering laser pulse.  As we have $a_1\approx286$, $ n_e/n_c\sim 100$ during the collision of the accelerated plasma with the scattering laser pulse (see Figs.~5(a1) and (b1) in the revised manuscript),   and $l_{RT}\sim\lambda_1$ is the charge separation in the transverse direction in the RT-instability, we can conclude that $E_1'\gg E'_{RT}$, and consequently, the electron dynamics  and related RR effects are mostly governed by the scattering laser field.

Moreover, we can estimate the effect of the scattering laser pulse on the plasma instability. When the RPA plasma interacts with the scattering laser pulse, the laser frequency $\omega_1$ is Doppler up-shifted in the rest frame of electron, $\omega_1'\approx \gamma_e (1 + \beta_e)\omega_1$, and the plasma frequency $\omega_{pe}'\approx \sqrt{4\pi n_e'e^2/(m a_1)}=\sqrt{4\pi n_ee^2/(\gamma_e m a_1)}$  due to the relativistic mass shift. Consequently, in the rest frame of electron $\frac{\omega_1'}{\omega_{pe}'} \approx  \gamma_e^{3/2}(1+\beta_e)\sqrt{a_1\frac{n_c}{n_e}} \gg 1$. The plasma thickness in the lab frame is defined as $l_p$, and consequently,  is $l_p\gamma_e$ in the rest frame of electron. A dimensionless parameter is defined, see Ref.~\cite{Vshivkov_1998}, in the rest frame of electron: 
\begin{eqnarray}
\epsilon_0= \frac{\omega_{pe}'^2 l_{p} \gamma_e}{2 \omega_1' c}\approx\frac{\pi n_e l_p}{n_c \lambda_1 \gamma_e (1+\beta_e)a_1}\ll a_1,\nonumber
\end{eqnarray}
indicating that the plasma is transparent for the scattering laser.  The latter precludes  the effective radiation pressure on the plasma, and inhibits the RT-like instability in the rear surface of plasma. Thus, we can conclude that the RT-like instability cannot disturb significantly the considered RR effects.

\section{Conclusion}\label{Sec5}

We have investigated the signature of SEs of photon emissions on the proton energy spectra in the RPA scheme, when the interaction of an ultrarelativistic plasma accelerated via RPA head-on collides with an ultraintense scattering laser pulse in the QRDR. A scattering laser pulse with judiciously chosen parameters  can significantly compress the proton spectra. We simulated the interaction with two methods, both including RR, however the MLL model neglects SEs, while the MC model fully takes into account them. By comparing the simulations of the MLL and MC models, we find that the SEs  induce a significant extra shift of the proton energy spectra, which is experimentally detectable
and can serve as a signature of SEs  with soon achievable ultraintense laser facilities, e.g., the ELI \cite{ELI}.

\section{acknowledgment}

This work is in part funded by the National Natural Science Foundation of China (Grants Nos. 11874295, 11804269, U1532263, 11875219), the  Science Challenge Project of China (No. TZ2016099), and the National Key Research and Development Program of China (Grant No. 2018YFA0404801), and  has been performed within the framework of the project High Field Initiative (CZ.02.1.01/0.0/0.0/15 003/0000449) from European Regional Development Fund.

\bibliography{refs}
\end{document}